\documentclass[aps,prl,twocolumn,groupedaddress]{revtex4}

\usepackage{graphicx}

\newcommand{\br}{{\bf r}}

\newcommand{\partder}[2]{\frac{\partial{#1}}{\partial{#2}}}

\newcommand{\be}{\begin{equation}}
\newcommand{\ee}{\end{equation}}

\begin{document}

\title{Splitting times of doubly quantized vortices in dilute Bose-Einstein condensates}
\author{J. A. M.\ Huhtam\"aki,$^1$ M.\ M\"ott\"onen,$^2$ T.\ Isoshima,$^3$ V.\
  Pietil\"a,$^1$ and S.\ M.\ M.\ Virtanen$^1$}

\affiliation{$^1$Laboratory of Physics, Helsinki University of
  Technology, P.O.\ Box 4100, FI-02015 TKK, Finland}
\affiliation{$^2$Low Temperature Laboratory, Helsinki University of
  Technology, P.O. Box 3500,
  FI-02015 TKK, Finland}
\affiliation{$^3$Institute of Physics, Academia Sinica,  Nankang, Taipei 11529, Taiwan}

\date{\today}

\begin{abstract}
Recently, the splitting of a topologically created doubly quantized vortex
into two singly quantized vortices was experimentally investigated in dilute
atomic cigar-shaped Bose-Einstein condensates~[Y.~Shin {\it et~al.},
Phys.\ Rev.\ Lett.\ \textbf{93}, 160406 (2004)]. In particular, the 
dependency of the splitting time on the peak particle density was
studied. We present results of theoretical simulations which closely mimic the 
experimental set-up. Contrary to previous theoretical studies, claiming that thermal
excitations are the essential mechanism in initiating the splitting, 
we show that the combination of gravitational sag and time
dependency of the trapping potential alone suffices to split the doubly 
quantized vortex in time scales which are in good
agreement with the experiments. We also study the dynamics of the resulting
singly quantized vortices which typically intertwine---especially, a
peculiar vortex chain structure appears for certain parameter values.
\end{abstract}


\maketitle

Bose-Einstein condensation is associated with long-range phase coherence 
among the condensed particles. Due to this coherence, the dynamics of
pure scalar condensates can be accurately described using only a single complex
valued order parameter. The fact that the condensate current density is
proportional to the phase gradient of the order parameter implies vorticity
to be quantized: vortices are topological defects in the order 
parameter field with a quantum number multiple~$\kappa$ of $2\pi$ phase 
winding about the vortex line.

Especially, properties of quantized vortices in gaseous atomic 
Bose-Einstein condensates (BECs) have been intensively investigated
during the recent years, both experimentally and theoretically, see
Ref.~\cite{rev} for a review. Since the angular momentum
associated with a vortex is roughly proportional to the vorticity
quantum number~$\kappa$ but the energy
to its square, multiquantum vortices are typically energetically unfavourable.
Consequently, the usual experimental methods to create vortices have 
yielded only singly quantized vortices or clusters of them~\cite{exp}. However,
using the topological phase engineering method originally suggested
by Nakahara {\em et al.}~\cite{Nakahara},
first two- and four-quantum vortices in dilute atomic
BECs have been realized \cite{Leanhardt}. This method utilizes
the hyperfine spin degrees of freedom of the order parameter, but finally
produces a scalar condensate containing a multiquantum vortex.

Theoretical analysis has revealed that in addition to multiquantum
vortices being energetically unfavourable in harmonic traps, they are also in general
dynamically unstable~\cite{Mikko,Pu}. This dynamical instability implies that
multiquantum vortices tend to split into singly quantized ones even 
in the absence of dissipation, {\it i.e.}, in pure condensates without
noticeable thermal gas component. The splitting dynamics is an interesting
issue, because the energy and angular momentum released from
the multiquantum vortex has to be redistributed in the system. The dynamics of the splitting of doubly
quantized vortices created in the experiments reported in Ref.~\cite{Leanhardt} 
was theoretically studied in Ref.~\cite{Mikko}, and two major observations 
were made: the initial dynamics in a cigar-shaped condensate can be 
modelled to some extent using only an effective two-dimensional local density
analysis, and proper three-dimensional computations showed 
that the two vortices separating out from the doubly quantized 
vortex usually intertwine strongly as they split.

Recently, the splitting of doubly quantized vortices was experimentally
observed and the splitting time, {\it i.e.}, the time interval between 
the creation of the vortex and the point when two separable vortex cores were observed, 
was measured as a function of the peak condensate density~\cite{Shin}. 
According to Ref.~\cite{Shin}, the temperature was low enough for
one to be able to neglect the effects of thermal atoms and dissipation.
The experimental results verified that the doubly quantized vortex 
splits into two singly quantized ones. However, since only particle 
densities averaged over a short slice 
in the longitudal direction of the condensate 
were measured, no intertwining of the vortices was observed.

In this Letter, we directly model the experiments reported in 
Ref.~\cite{Shin}, and compare the theoretical and experimental results.
We solve the full three-dimensional dynamics of the condensate using
the Gross-Pitaevskii (GP) equation with time-dependent trapping potential
combined with gravitational potential, closely corresponding to the 
experiments. From the condensate density profiles, we determine the
splitting time as a function of the peak condensate density, and 
analyse the effects of vortex intertwining. Contrary to previous 
theoretical results presented in Ref.~\cite{Gawryluk}, we find that 
the gravitational and the time-dependent trapping 
potentials together break the rotational symmetry 
and initiate the splitting process strongly enough to alone yield
splitting times in good agreement with experiments. Thus the
effect of thermal excitations is not relevant in modelling
the experiments.

At low enough temperatures, the
effect of the thermal gas component can be neglected
and the entire gaseous many-particle system of trapped atoms can be
described by the condensate order parameter $\psi(\br,t)$. Dynamics of this
dilute condensate is governed by the GP equation
\be\label{gp}
i\hbar\partder{}{t}\psi(\br,t)=\left[-\frac{\hbar^2}{2m}\nabla^2+V(\br,t)+g|\psi(\br,t)|^2\right] \psi(\br,t),
\ee
where $V(\br,t)$ is the external potential, $m$ the mass of the atoms,
and the strength of the interactions is governed by the parameter
$g=4\pi\hbar^2 a/m$ expressed in terms of the $s$-wave scattering length $a$. The
order parameter is normalized according to the total number of atoms as
$\int |\psi|^2 \textrm{d}\br = N$. The stationary state solutions of the GP
equation with eigenvalue $\mu$
are of the form $\psi(\br,t)=e^{-i\mu t/\hbar}\psi(\br)$, and the excitation
spectrum of these states can be solved from the Bogoliubov
equations. The excitation spectrum determines the stability properties of
the state: Existence of modes with negative energy but positive norm imply the corresponding stationary state to be energetically
unstable, {\it i.e.}, in the presence of dissipation and small perturbations the
state will decay. Furthermore, existence of modes with non-real eigenfrequencies
implies the state to be dynamically unstable, {\it i.e.}, infinitesimal
perturbations may grow exponentially in time, and thus the stationary state may
decay even in the absence of dissipation.

Examples of these kind of states are the axisymmetric vortex states of the form
$\psi(\br)=e^{i2\pi\kappa\theta}\psi(r,z)$, where $(\theta,r,z)$ denote the
cylindrical coordinates and $\kappa\neq 0$. These states are
energetically unstable in harmonic traps since vortices tend to spiral out of the condensate
due to the buoyancy force. In addition, for quantum numbers
$|\kappa|\geq 2$ there typically exists excitations with non-real eigenfrequencies which imply
the corresponding states to be dynamically unstable and to split into singly
quantized vortices. Especially, the dynamical instability of doubly quantized vortices has been demonstrated both
theoretically~\cite{Pu,Mikko} and experimentally~\cite{Shin}.

In the experiments reported in Ref.~\cite{Shin}, Y. Shin {\it et al.} created a doubly quantized vortex into a
dilute BEC of $^{23}$Na atoms confined in a Ioffe-Pritchard trap by reversing the axial bias
field $B_z$ linearly in 12 ms. The thermal gas fraction was not discernible in the
experiments, and hence dissipation was presumably extremely weak.

We have modelled these experiments by computing the time evolution of the
condensate using the GP equation. Starting from energy minimized axisymmetric doubly quantized
vortex states corresponding to different values of the density parameter
$an_{z}=a\int|\psi(x,y,0)|^2 \textrm{d}x\textrm{d}y$ (at the center of the
condensate $z=0$), we calculate the
condensate dynamics in the time-dependent
external potential corresponding to the Ioffe-Pritchard trap and gravity. Taking
the gravity into account turns out to be crucial, since only the combined
potential breaks the rotational symmetry about the vortex axis and is shown to initiate
the vortex splitting process. In the computations, we used finite-difference
discretization. The initial states were found using imaginary time
integration, and the time evolution was calculated by a split operator method.

In a Ioffe-Pritchard trap, the square of the total magnetic field strength to the second order in the
radial and axial coordinates is given by~\cite{PethickSmith}
\be
\label{eq2}
B^2(t)=C^2 r^2+B_z(t)A[z^2-\frac{1}{2} r^2]+B^2_z(t),
\ee
where $A$ determines the curvature of the
initial axial field, $C$ characterizes the strength of the field generated
by the Ioffe bars and $B_z$ is the axial bias field. Including gravity, the external
potential for the weak-field seeking state of spin-1 condensate reads
\be
V(\br,t)=-g_{\rm L} \mu_{\rm B} B(t)+G m x,
\ee
where $g_{\rm L}=-1/4$ and $G$ is the gravitational acceleration. At $t=0$,  the term $B²_z(0)$
dominates in~Eq.\ (\ref{eq2}) in the vicinity of the center
of the trap, and hence the total potential is approximately harmonic in the
condensate region. 
The initial potential is of the form
\be
V(\br,0) \approx \frac{1}{2}m \omega_z^2 z^2 + \frac{1}{2}m \omega_{r}^2 [y^2+(x-x_0(0))^2]+\textrm{const.}
\ee
According to the experiments, we choose the parameters $A$ and $C$ such
that the axial and radial trapping frequencies are $\omega_z=12~\textrm{Hz}$ and
$\omega_{r}=220~\textrm{Hz}$, respectively. In addition, $x_0(t)=Gm|B_z(t)|/(g_{\rm L} \mu_{\rm B}[C^2-B_z(t)A/2])$ is the location of the minimum of
$V(\br,t)$. The bias field $B_z$ vanishes at $t=6~\textrm{ms}$, and hence the radial trapping potential is
linear with respect to $r$. As $B_z(t)$ passes through zero, the sign of
$A$ is changed in order to keep the axial field confining. 
After the reversal of $B_z$ at $t=12~\textrm{ms}$, the potential assumes its initial form.

\begin{figure}
\includegraphics[width=7.0cm]{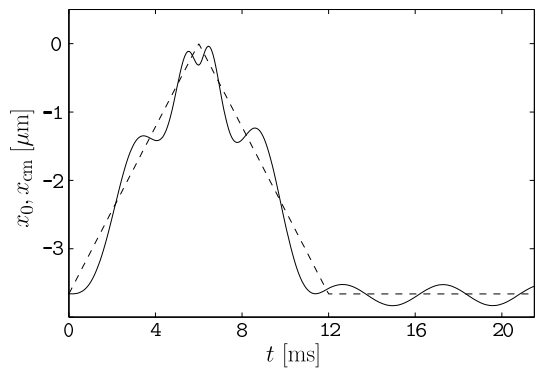}
\caption{\label{potminandmkp} Location $x_0(t)$ of the minimum of the total
  potential $V(\br,t)$ (dashed line) and the location $x_{\rm{cm}}(t)$ of the center of mass of the condenstate
  (solid line). Due to inertia, the condensate lags behind the center of the
  potential, which breaks the rotational symmetry of the system. This
  perturbation initiates the splitting of the doubly quantized vortex.}
\end{figure}

\begin{figure}
\includegraphics[width=8.5cm]{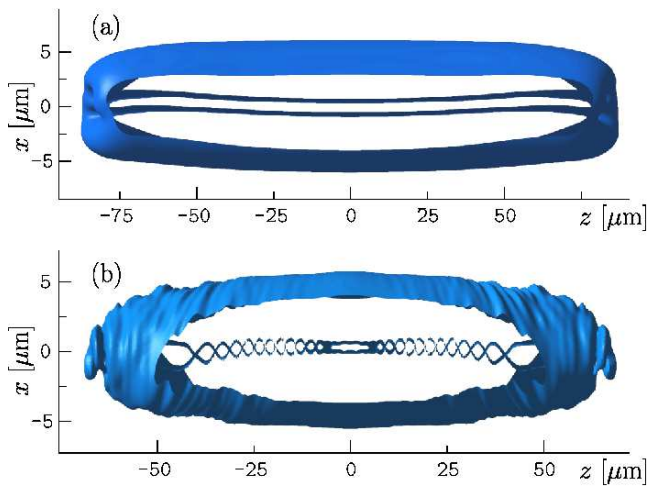}
\caption{\label{smurfs} (color online) Isosurfaces of condensate densities for (a)
  $an_{z}=2.56$ at time $t=10.9~\textrm{ms}$ and (b)
  $an_{z}=14.4$ at $t=39.8~\textrm{ms}$. The values defining the isosurfaces are of the order of $1$\% of
  the maximal density.}
\end{figure}

\begin{figure}
\includegraphics[width=8.5cm]{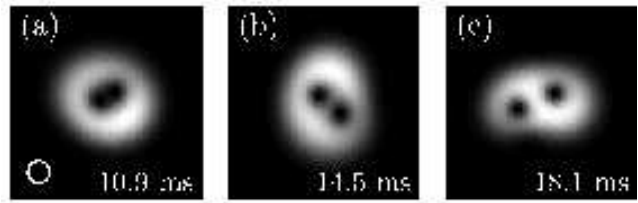}
\caption{\label{ufos}Density profiles $\bar{n}(x,y)=\int_{-15~\mu \rm{m}}^{15~\mu
   \rm{m}}|\psi(x,y,z)|^2 \textrm{d}z$ of a condensate with $an_{z}=2.56$ at three different times $t$. The splitting time of the doubly quantized
  vortex is determined by comparing the separation of the local minima in the
  density profiles to the diameter of a single quantum vortex represented by
  the white circle corresponding to a diameter of $1.8~\mu \rm{m}$ in (a).
  The field of view in these figures is $13.5~\mu \textrm{m} \times 13.5~\mu \textrm{m}$.}
\end{figure}

\begin{figure}
\includegraphics[width=7cm]{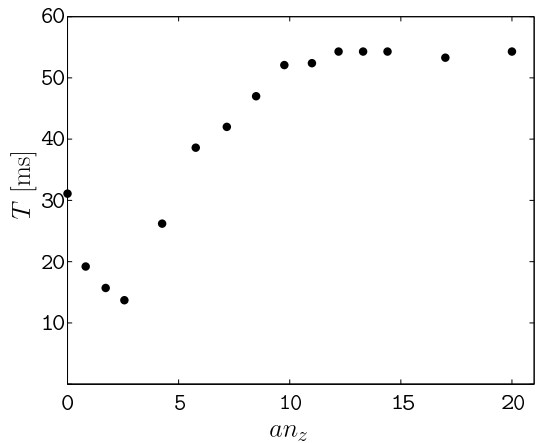}
\caption{\label{splittingtimes} The splitting time $T$ of a doubly quantized
  vortex as a function of $an_{z}$. The splitting time
  decreases rapidly as interactions are introduced in the system attaining its
  minimum value around $an_{z} \approx 3$. As the interaction strength is
  increased further, $T$ increases due to stable region near the center of the
  condensate. The splitting time saturates roughly to $T \approx 55~\textrm{ms}$ for
  interaction strengths $an_{z} > 10$.}
\end{figure}

\begin{figure}
\includegraphics[width=8.5cm]{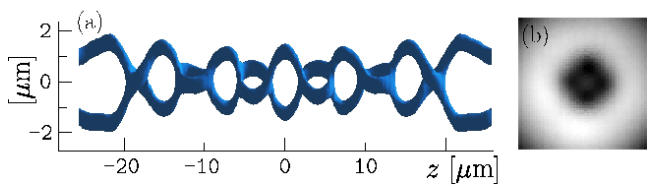}
\caption{\label{vortexchain} (color online) (a) Isosurface plot of the vortex chain structure
  formed in a condensate with $an_{z}=11.0$ at $t=54.3~\textrm{ms}$. The
  value defining the isosurface is of the order of $10$\% of the maximal density. (b) The
  corresponding density profile $\bar{n}(x,y)$ shows four minima instead
  of two. The field of view in (b) is $6.8~\mu \textrm{m} \times 6.8~\mu \textrm{m}$.}
\end{figure}

The main results of our simulations are presented in Figs.\ \ref{potminandmkp} -- \ref{vortexchain}.
Figure~\ref{potminandmkp} shows the location $x_0(t)$ of the minimum
of $V(\br,t)$ and the computed location of the center of mass of the condensate
as functions of time. Since the condensate slightly lags behind the
time-dependent center of the potential, the rotational symmetry of the initial
condensate state is broken. This perturbation also excites
the dipole mode of the condensate, which is manifested by the oscillatory
behavior of the center of mass after the potential has returned to its initial
form. However, the amplitude of the dipole oscillation is less than 3\% of the diameter of the condensate.

In the effectively two-dimensional case, for which $\omega_z=0$ and the condensate
is homogenous in the $z$-direction, the imaginary part of
the frequency corresponding to the eigenmode responsible for the dynamical instability assumes a quasi-periodic form as function of
$an_z$~\cite{Mikko,Pu}, and vanishes in some regions of the parameter
$an_z$. Also in three-dimensional cigar-shaped condensates,
regions where splitting is significantly faster or slower are clearly
distinguishable. In Fig.~\ref{smurfs}(a), the isosurface of a condensate with
$an_{z}=2.56$ is plotted at $t = 10.9~\textrm{ms}$. In this case, the splitting
process is initiated along the whole length of the condensate, to be
contrasted to the case $an_{z}=14.4$ at $t = 39.8~\textrm{ms}$ depicted
in Fig.~\ref{smurfs}(b), in which the splitting initiates only from the ends
and center of the condensate. Since the precession frequency of the
vortices depends on their distance, intertwining is observed in cases which
contain such stable regions. Note also the distinctive surface mode
excitations due to the additional energy and angular momentum released in the
splitting process.

We determine the splitting times $T$ from the density profiles
$\bar{n}(x,y)=\int_{-15~\mu \rm{m}}^{15~\mu \rm{m}}|\psi(x,y,z)|^2 \textrm{d}z$, mimicking
the tomographic imaging technique used in the experiments~\cite{Shin}. 
Roughly, when two minima are observed in the density profile
with their separation exceeding the diameter of a singly quantized
vortex, the doubly quantized vortex was considered split in the experiments.
We define the diameter of a single quantum vortex as twice the
distance from the vortex center to $75 \%$ of the maximum value in the
density profile $\bar{n}(x,y)$ of the stationary single quantum vortex state. In
Fig.~\ref{ufos}, the density profile $\bar{n}(x,y)$ is plotted at $t=10.9,
14.5~\textrm{and}~18.1~\textrm{ms}$ for $an_{z}=2.56$. From the isosurface
plot in Fig.~\ref{smurfs}(a), one might argue that the vortex has already
split at $t=10.9$ ms, but from the density plot in Fig.~\ref{ufos}(a) no such
conclusions can be made. In Fig.~\ref{ufos}(b), the distance of the local density
minima already exceeds the single quantum vortex diameter and a detailed
analysis yields $T=13.7$ ms for the splitting time.

Figure \ref{splittingtimes} shows the splitting times for
values of $an_{z}$ ranging from $0$ to $20$ corresponding up to
$N= 8 \times 10^5$ ${}^{23}$Na atoms. A finite splitting time $T$ is obtained also in the
noninteracting case, which concurs with the observation that persistent
stationary currents exist in the noninteracting case only if the
spectrum of the Hamiltonian is degenerate \cite{GarciaRipoll}. For $an_z < 3$,
$T$ decreases with increasing interaction strength. This effect is also visible
in the recent theoretical study of Ref.~\cite{Gawryluk}, but it is not clearly
observable in the experimental results~\cite{Shin}. The splitting time attains its minimum at
$an_{z}~\approx~3$, which is in good agreement with the Bogoliubov
eigenvalue spectrum analysis accomplished in Refs.~\cite{Mikko,Pu}. For $an_{z}> 10$, $T$ saturates to a finite value, which is
in fair agreement with the experimental results \cite{Shin}, although
the saturation value is roughly $20\%$ larger in the experiments. The overall
form of the computed splitting times as a function of $an_z$ agrees well with the experimental data, and the
agreement is surprisingly good even for the quantitative results. Note that the computed splitting times are generally somewhat
smaller than the measured ones, implying that the time dependency of the
external potential alone
gives sufficiently strong impetus for the splitting, and the additional contribution of
thermal fluctuations is not needed.

A few points should be taken into account when comparing the computed
splitting times to experimental values. Firstly, our simulations neglect
the multicomponent nature of the condensate when the multiquantum vortex is
created by reversing the bias field $B_z$. In our simulations, we
begin with the multiquantum vortex state already at $t=0$. 
Effectively, this difference can roughly be assumed to shorten the computed lifetimes
of the multiquantum vortices by half of the reversing time, {\it i.e.}, by $6$ ms.
However, this difference is presumably more than compensated by the fact that
in the simulations we start counting the splitting time from the beginning of the perturbation,
whereas in the experiments the time is counted from the end of the
perturbation. Secondly, in the experiments the condensate is let to expand
freely for $15~\textrm{ms}$ before imaging, and in our results possible changes in the relative
vortex separation during this expansion period are neglected.

For values near $an_{z} \approx 10$, where there is a stable region at
the center of the condensate, a peculiar linked chain-type structure for
the intertwined vortices was observed, see Fig.~\ref{vortexchain}(a).
The $z$-integrated density profile observed with the tomographic imaging
technique is shown in Fig.~\ref{vortexchain}(b). In this case, there are four
distinct minima in the density profile instead of the usual two. This effect could explain
the observation interpreted as crossing of the vortex lines in the experiments.

In conclusion, we have calculated the splitting times of doubly quantized
vortices as a function of peak particle density by solving the time-dependent Gross-Pitaevskii equation
numerically, mimicking closely the experiments reported in Ref.~\cite{Shin}. The
results are in good agreement with the experimental data, confirming that thermal
fluctuations are not required to explain the measured splitting times. Instead, the main impetus for the rotational symmetry
breaking and splitting is the combination of the gravitational and
time-dependent trapping potentials.

The authors thank CSC, the Finnish IT Center for Science, for
computational resources. J. H. and M. M. acknowledge the Finnish Cultural Foundation
for financial support. Y. Shin is appreciated for helpful discussions.

\end{document}